\documentclass[12pt]{article}
\usepackage{setspace}

\usepackage{amsmath}
\usepackage{jtb}

\ifx\pdftexversion\undefined
  \usepackage[dvips]{graphicx}
  \usepackage{pstricks}
  \usepackage{pst-node}
\else
  \usepackage[pdftex]{graphicx}
\fi

\newcommand{\micromol}[0]{$\mu\mbox{M}$}
\newcommand{\millimol}[0]{$\mbox{mM}$}

\begin{document}

\begin{center}
\noindent
{\Large\bf Incorporating expression data in metabolic modeling: a case
  study of lactate dehydrogenase}
 
\vspace{0.4cm}
\noindent
{\bf Joshua Downer$^{\dagger}$, Joel R. Sevinsky$^{\ast}$, Natalie G.
  Ahn$^{\ast}$, Katheryn A. Resing$^{\ast}$, M. D. Betterton$^{\dagger}$}
\end{center}

\begin{center}
\noindent
$\dagger$Department of Applied Mathematics, University of Colorado at
Boulder, 526 UCB, CO 80309-0526, USA, and \\ $\ast$Department of
Chemistry and Biochemistry, University of Colorado at Boulder,
215 UCB, CO 80309-0215, U.S.A.
 
\vspace{0.6cm}
\noindent
 
\end{center}
\vspace{0.6cm}
\noindent
Address correspondence to: M. D. Betterton, University of Colorado at
Boulder, Department of Applied Mathematics, 526 UCB, CO 80309-0526,
U.S.A., Email: mdb@colorado.edu, Phone: 303-735-6135.
 

 \Section{Abstract}
 
 Integrating biological information from different sources to
 understand cellular processes is an important problem in systems
 biology. We use data from mRNA expression arrays and chemical
 kinetics to formulate a metabolic model relevant to K562
 erythroleukemia cells. MAP kinase pathway activation alters the
 expression of metabolic enzymes in K562 cells. Our array data show
 changes in expression of lactate dehydrogenase (LDH) isoforms after
 treatment with phorbol 12-myristate 13-acetate (PMA), which activates
 MAP kinase signaling.  We model the change in lactate production
 which occurs when the MAP kinase pathway is activated, using a
 non-equilibrium, chemical-kinetic model of homolactic fermentation.
 In particular, we examine the role of LDH isoforms, which catalyze
 the conversion of pyruvate to lactate.  Changes in the isoform ratio
 are not the primary determinant of the production of lactate.
 Rather, the total concentration of LDH controls the lactate
 concentration.


 \Section{Keywords}
 
 metabolic modeling, MAP kinase, cell signaling, enzyme kinetics,
 expression data, lactate dehydrogenase

\Section{Introduction}

Modeling of cellular metabolism has a long history of important
contributions to biology. Approaches include kinetic modeling,
metabolic control analysis, flux balance analysis, and metabolic
network analysis \cite{steph98}.  A new era of research in metabolism
is now possible, because large-scale expression studies can determine
levels of many metabolites \cite{goodac04,fan04} and metabolic enzymes
\cite{ferea99,kal99}. In this paper we use metabolic enzyme expression
data to guide metabolic modeling, with a focus on small but
significant changes in mRNA abundance. Our goal is to understand how
biologically realistic changes in mRNA abundance of metabolic enzymes
affect cellular metabolism. Previous work integrating expression data
with metabolic modeling has been done in yeast \cite{akess04}, but
not, to our knowledge, in mammalian systems.

We focus on glycolysis, an essential ATP-producing metabolic pathway.
The initial reactions of glycolysis break down glucose into pyruvate.
Pyruvate can feed into either the citric acid cycle (aerobic
metabolism) or homolactic fermentation (anaerobic metabolism)
\cite{voet04}.  The reactions involving pyruvate therefore control
this important metabolic branch point. Homolactic fermentation is
catalyzed by lactate dehydrogenase (LDH) in a compulsory-order,
ternary reaction \cite{borgmann75}. LDH reversibly converts pyruvate
and NADH into lactate and NAD+.  The isozymes of LDH are tetramers
formed from two types of monomers (a third isoform is usually
germ-line specific, but can be expressed in cancers
\cite{koslowski02}). The two isoforms are labeled H (heart) and M
(muscle), and their ratio varies between cell types. The LDH isoform
ratio has been proposed to indicate the metabolic state of cells: it
is believed that the M isoform favors lactate production while the H
isoform favors pyruvate production
\cite{boyer63,stambaugh66,boyer75,voet04}. In this framework, the LDH
isoform ratio can serve as an indicator of the relative flux through
aerobic/anaerobic gycolytic pathways.

Here we use a mathematical model of homolactic fermentation to study
the connections between growth-factor signaling and metabolism. It has
been known since the work of Warburg that carcinogenesis is
accompanied by changes in cellular metabolism
\cite{stubb03,griff02,dang99}. In particular, tumors typically favor
anaerobic metabolism, resulting in higher lactate production relative
to non-cancerous cells \cite{walen04,newell93,warburg56}.  Although
inhibition of glycolysis can kill tumor cells \cite{munoz03}, the
connections between carcinogenesis and metabolic alterations are not
fully understood \cite{fan04}. However, intriguing connections between
metabolic enzymes and cancer have been demonstrated
\cite{kondoh05,kim04,mazur03}. In particular, LDH expression is
altered in many tumors \cite{walen04,unwin03,maekaw03} and cancer cell
models \cite{li04,karan02,lewis00}. High tumor LDH levels have been
shown to correlate with poor prognosis in lung cancer patients
\cite{koukour03}.

In this study, we focus on changes in LDH expression induced by the
mitogen-activated protein (MAP) kinase pathway.  The MAP kinase
cascade is important in cell growth, differentiation, and survival,
and alterations of MAP kinase signaling have been found in many
cancers \cite{lewis98}.  The signal is transduced by a series of
phosphorylation reactions: MAP kinase proteins phosphorylate and
thereby activate their downstream targets. The pathway includes the
MAP kinase proteins ERK 1 and 2 and their upstream activators, the MAP
kinase kinases MKK 1 and 2.  Recent work has found connections between
MAP kinase signaling and metabolism. For example, increased expression
of LDH-H in human tumors may occur in part because the transcription
factor MYC, a downstream target of the MAP kinase pathway,
transcriptionally up-regulates the LDH-H gene \cite{jungm98,shim97}.
Other genes involved in glycolysis are also affected by MYC
\cite{osthus00}.  Activation of the MAP kinase pathway has been shown
to increase LDH activity, glucose uptake, and lactate production
\cite{riera03,papas99}.

\begin{figure}[t] \centering
\includegraphics[width=5in]{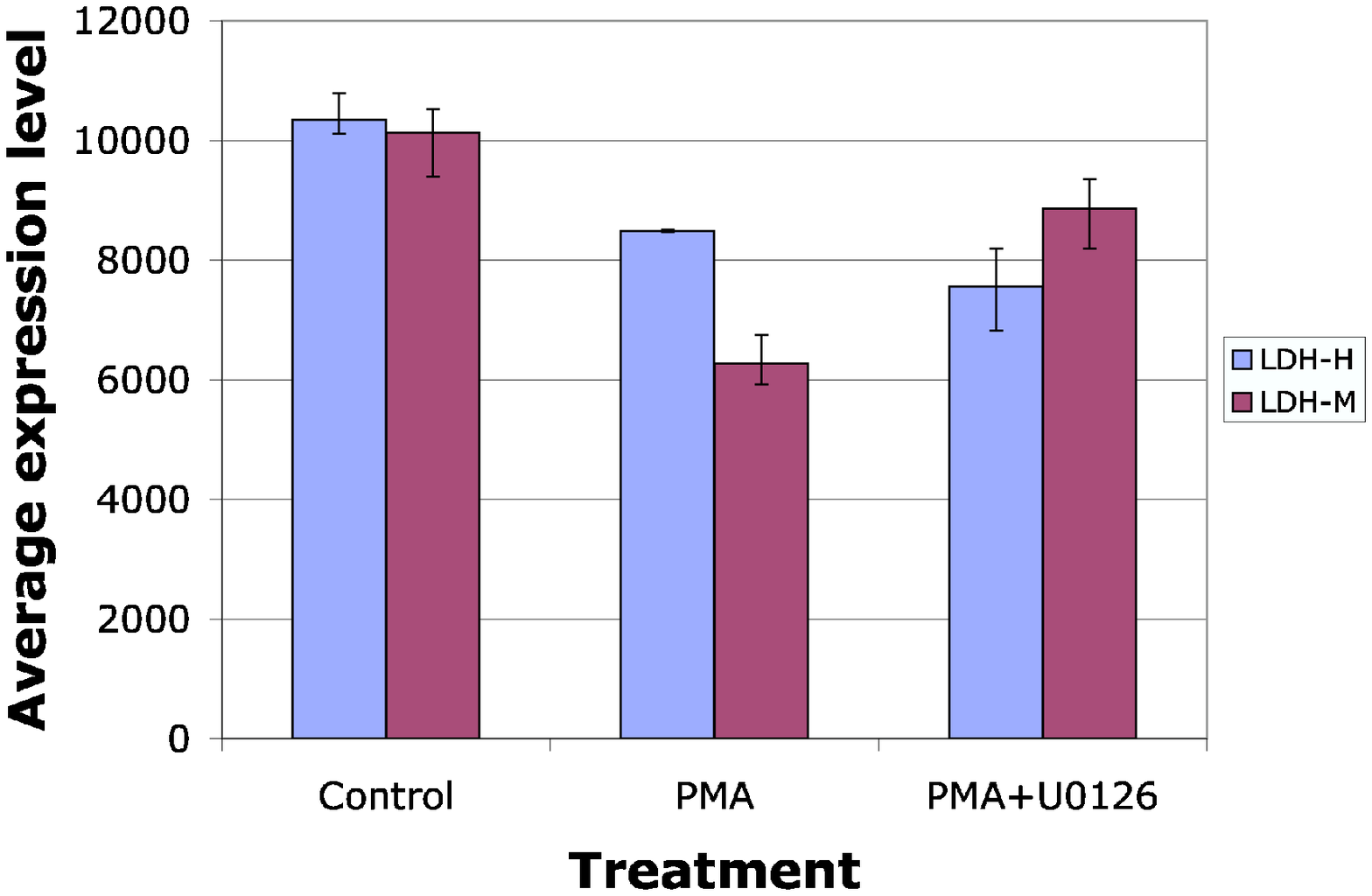}
\noindent
\caption[Changes in LDH-H and LDH-M mRNA expression after treatment
with PMA and PMA+U0126. PMA treatment activates the MAP kinase
pathway, while U0126 is a downstream MAP kinase inhibitor. The data
were obtained from experiments performed on Affymetrix gene chips in
triplicate (see text).  Each data point is the average of three
measurements and the error bars represent the maximum and minimum
values. Treatment with PMA reduces the abundance of both LDH isoforms,
and the M isoform shows a larger reduction.  Treatment with PMA+U0126
reduces the abundance of both isoforms (relative to control), but the
H isoform shows a larger reduction. The expression levels of LDH-M are
significantly different among all of the treatments ($p < 0.07$). The
expression levels of LDH-H for the PMA and PMA+U0126 treatments are
not significantly different from one another, but they are both
significantly different from the expression level of LDH-H in the
control ($p < 0.02$).]{Changes in LDH-H and LDH-M mRNA expression
  after treatment with PMA and PMA+U0126. PMA treatment activates the
  MAP kinase pathway, while U0126 is a downstream MAP kinase
  inhibitor. The data were obtained from experiments performed on
  Affymetrix gene chips in triplicate (see text).  Each data point is
  the average of three measurements and the error bars represent the
  maximum and minimum values. Treatment with PMA reduces the abundance
  of both LDH isoforms, and the M isoform shows a larger reduction.
  Treatment with PMA+U0126 reduces the abundance of both isoforms
  (relative to control), but the H isoform shows a larger reduction.
  The expression levels of LDH-M are significantly different among all
  of the treatments ($p < 0.07$). The expression levels of LDH-H for
  the PMA and PMA+U0126 treatments are not significantly different
  from one another, but they are both significantly different from the
  expression level of LDH-H in the control ($p < 0.02$).}
\label{fig:affy_hist}
\end{figure}

We studied mRNA expression in K562 erythroleukemia cells, a cell line
used as a model for leukemia. In our experiments, MAP kinase signaling
was either (\textit{i}) activated with phorbol 12-myristate 13-acetate
(PMA) or (\textit{ii}) simultaneously activated with PMA and inhibited
with U0126, a specific MKK inhibitor. We found small but reproducible
changes in the expression of LDH isoforms in response to MAP kinase
pathway activation (figure \ref{fig:affy_hist}), with no significant
changes in other enzymes that catalyze reactions involving pyruvate.
This result suggests that activating the MAP kinase pathway alters the
relative flux through aerobic and anaerobic glycolysis in these cells.
We chose to model the expected changes in cellular lactate prodution
to better understand the connections between signaling and metabolism.
We hypothesized that the LDH isoform ratio plays an important role in
determining cellular lactate levels, as suggested previously
\cite{riera03,dang99}.

We formulated a chemical-kinetic model of homolactic fermentation
based on \textit{in vitro} biochemistry \cite{borgmann75}. Our goal
was to determine how changes in the LDH isoform ratio alter the amount
of lactate produced by K562 cells. We used the experimentally
determined abundance changes as model inputs. The model describes the
mass-action kinetics of homolactic fermentation. We included metabolic
flux terms in the model to describe the connection between homolactic
fermentation and the larger metabolic network of the cell.  The
metabolic flux is a constant rate of production/consumption of a
metabolite through other reactions or transport.  Several model
inputs---the steady-state concentrations of pyruvate, NADH, and
NAD+---have not been measured in K562 cells.  Therefore we validated
our results with a robustness analysis \cite{dassow00,barkai97}.
 
We present several unexpected findings. In a preliminary analysis, we
examined the behavior of each isoform individually.  Our results predict
that LDH-H produces a larger steady-state lactate concentration than
an equivalent amount of LDH-M under typical cellular conditions. This
result is surprising because it disagrees with the statement, often
found in the literature, that the M isoform favors lactate production
while the H isoform favors pyruvate production
\cite{boyer63,stambaugh66,boyer75,voet04}. We discuss the reason for
this difference and explain why our results are more applicable
\textit{in vivo}.
 
Second, we predict a decrease in the steady-state lactate
concentration when the LDH isoform abundance shifts from control to
PMA-treated levels.  This result means that the H:M isoform ratio
alone does not control the lactate concentration. After PMA treatment
the ratio of LDH-H to LDH-M changes from 1.02 to 1.35 in our
experiments.  According to our single-isoform model, an increasing
isoform ratio should lead to an {\em increase} in lactate
concentration.  This finding led us to consider separately how the
isoform ratio and the total abundance of LDH control the lactate
concentration. We demonstrate that while the isoform ratio does affect
the production of lactate, the experimentally determined total LDH
abundance change plays a larger role in determining the lactate
concentration.

\Section{Methods}
\label{methods}

\SubSection{Cell extraction and microarray analysis}
\label{methods:affy}

K562 erythroleukemia cells were grown in suspension in 10\% FBS/RPMI
and treated with 10 nM phorbol 12-myristate 13-acetate (PMA) and 20
\micromol~U0126 (Promega) as described previously \cite{sevinsky04}.
Cells (7 x 105) were washed twice in ice cold phosphate buffered
saline, 1 mM EDTA, 1 mM EGTA, and total RNA was isolated by TRIzol
extraction (Invitrogen).  First and second strand cDNA synthesis,
\textit{in vitro} transcription of biotin-labeled cRNA, and
fragmentation were carried out following standard protocols from
Affymetrix. Probes were hybridized onto U133 2.0 Plus GeneChips
(Affymetrix) and processed at the UCHSC Cancer Center Microarray core
facility.  Datasets were corrected for background and normalized using
RMA Express software.  Each condition was analyzed in three
independent experiments (figure \ref{fig:affy_hist}).

\SubSection{Model}
\label{methods:model}

We used a chemical-kinetic model to analyze the effect of isoform
switching on the non-equilibrium steady state of homolactic
fermentation. We assume that the metabolite and enzyme species are
homogeneously distributed in the cytosol. This leads to a set of 12
mass-action ordinary differential equations that describe the time
evolution of metabolite and enzyme concentrations.  Four equations
govern the metabolites and four equations govern each of the LDH
isoforms and their related complexes (equations~\ref{m1}-\ref{m3}).

\begin{figure}[t]
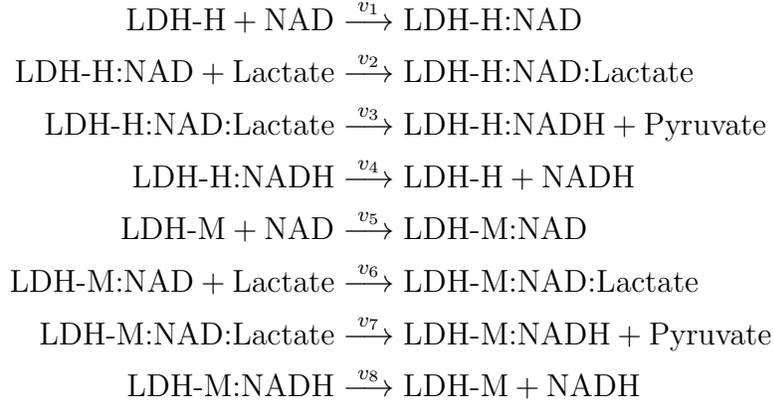
\centering
\begin{minipage}{0.3\textwidth}
\begin{align*}
\text{LDH-H} + \text{NAD} &\stackrel{v_1}{\longrightarrow} \text{LDH-H:NAD} \\
\text{LDH-H:NAD} + \text{Lactate} &\stackrel{v_2}{\longrightarrow} \text{LDH-H:NAD:Lactate} \\
\text{LDH-H:NAD:Lactate} &\stackrel{v_3}{\longrightarrow} \text{LDH-H:NADH} + \text{Pyruvate} \\
\text{LDH-H:NADH} &\stackrel{v_4}{\longrightarrow} \text{LDH-H} + \text{NADH} \\
\text{LDH-M} + \text{NAD} &\stackrel{v_5}{\longrightarrow} \text{LDH-M:NAD} \\
\text{LDH-M:NAD} + \text{Lactate} &\stackrel{v_6}{\longrightarrow} \text{LDH-M:NAD:Lactate} \\
\text{LDH-M:NAD:Lactate} &\stackrel{v_7}{\longrightarrow} \text{LDH-M:NADH} + \text{Pyruvate} \\
\text{LDH-M:NADH} &\stackrel{v_8}{\longrightarrow} \text{LDH-M} + \text{NADH}
\end{align*}
\end{minipage}
\caption[The elementary reactions of homolactic fermentation. Homolactic
fermentation is a compulsory-order, ternary reaction. All elementary
reactions are reversible.  The arrow in each elementary reaction
indicates the direction of a positive reaction rate.]{The elementary
  reactions of homolactic fermentation. Homolactic fermentation is a
  compulsory-order, ternary reaction. All elementary reactions are
  reversible.  The arrow in each elementary reaction indicates the
  direction of a positive reaction rate.}
\label{fig:schematic}
\end{figure}

Each elementary reaction in the model (figure~\ref{fig:schematic})
follows the law of mass action, which results in the following
reaction rates
\begin{subequations}
\label{m4}\begin{align}
v_1 & =  k_1 x_1 e_1 - k_{-1} e_2, \\
v_2 & =  k_2 x_2 e_2 - k_{-2} e_3, \\
v_3 & =  k_3 e_3 - k_{-3} y_1 e_4, \\
v_4 & =  k_4 e_4 - k_{-4} y_2 e_1, \\
v_5 & =  k_5 x_1 f_1 - k_{-5} f_2, \\
v_6 & =  k_6 x_2 f_2 - k_{-6} f_3, \\
v_7 & =  k_7 f_3 - k_{-7} y_1 f_4, \\
v_8 & =  k_8 f_4 - k_{-8} y_2 f_1,
\end{align}
\end{subequations}
where $x_1$, $x_2$, $y_1$, and $y_2$ are the concentrations of NAD+,
lactate, pyruvate, and NADH; $e_1$, $e_2$, $e_3$, and $e_4$ are the
concentrations of LDH-H, LDH-H:NAD+, LDH-H:NAD+:lactate, and
LDH-H:NADH; $f_1$, $f_2$, $f_3$, and $f_4$ are the concentrations of
LDH-M, LDH-M:NAD+, LDH-M:NAD+:lactate, and LDH-M:NADH. The values of
the kinetic rate constants are shown in figure \ref{fig:model}.

The equations describing the dynamics of the system can be written
compactly in terms of the reaction rates. The equations for the
metabolites are
\begin{subequations}
\label{m1}
\begin{align}
  x'_1 &= -v_1 - v_5 - \alpha_{1}, \label{eq1a} \\
  x'_2 &= -v_2 - v_6 - \alpha_{2}, \label{eq1b} \\
  y'_1 &= v_3 + v_7 + \alpha_{3}, \label{eq1c} \\
  y'_2 &= v_4 + v_8 + \alpha_{4}, \label{eq1d}
\end{align}
\end{subequations}
where $\alpha_1$ and $\alpha_2$ are the flux of NAD+ and lactate
out of the system, and $\alpha_3$ and $\alpha_4$ are the flux
of NADH and pyruvate into the system. 

LDH is an efficient catalyst which is thought to operate near
equilibrium under many conditions \cite{borgmann75}. However, the
metabolites in homolactic fermentation are continually consumed or
produced in other reactions or transported into and out of the cell.
This flux of metabolites means that the system is not in thermodynamic
equilibrium.  In contrast to an equilibrium system, the steady state
of a non-equilibrium reaction depends on the reaction mechanism and
the total concentration of the enzyme that catalyzes it. Here we do
not specify a mechanism for this metabolic flux but assume that each
metabolite is produced or consumed at a constant rate.  The constant
flux of each metabolite in the model is represented by the constant
terms $\alpha_{1}, \ldots, \alpha_{4}$ in equation (\ref{m1}).

The heart-isoform complexes obey the equations
\begin{subequations}
\label{m2}
\begin{align}
e'_1 &= v_4 - v_1 \label{eq2a}, \\
e'_2 &= v_1 - v_2 \label{eq2b}, \\
e'_3 &= v_2 - v_3 \label{eq2c}, \\
e'_4 &= v_3 - v_4 \label{eq2d},
\end{align}
\end{subequations}
and the muscle-isoform complexes obey the equations
\begin{subequations}
\label{m3}
\begin{align}
f'_1 &= v_8 - v_5 \label{eq3a}, \\
f'_2 &= v_5 - v_6 \label{eq3b}, \\
f'_3 &= v_6 - v_7 \label{eq3c}, \\
f'_4 &= v_7 - v_8 \label{eq3d}.
\end{align}
\end{subequations}
\par

We are primarily interested in the steady-state behavior of the model,
which occurs only when the fluxes are equal for all four metabolites.
In other words, a steady state exists if and only if $\alpha_{1} =
\alpha_{2} = \alpha_{3} = \alpha_{4} = \alpha$, where $\alpha$ is the
constant metabolic flux of the model.
\par  

\begin{figure} \centering
\begin{minipage}{0.8\textwidth}
\begin{minipage}{0.45\textwidth}
\begin{eqnarray*}
k_1 & = & 1.45 \times 10^6 \, {\rm M^{-1} s^{-1}} \\
k_2 & = & 2.06 \times 10^5 \, {\rm M^{-1} s^{-1}} \\
k_3 & = & 3.29 \times 10^4 \, {\rm s^{-1}} \\
k_4 & = & 4.33 \times 10^2 \, {\rm s^{-1}} \\
k_5 & = & 7.50 \times 10^5 \, {\rm M^{-1} s^{-1}} \\
k_6 & = & 4.10 \times 10^4 \, {\rm M^{-1} s^{-1}} \\
k_7 & = & 1.51 \times 10^4 \, {\rm s^{-1}} \\
k_8 & = & 6.65 \times 10^2 \, {\rm s^{-1}}
\end{eqnarray*}
\end{minipage}
\hspace{0.15\textwidth}
\begin{minipage}{0.45\textwidth}
\begin{eqnarray*}
k_{-1} & = & 1.88 \times 10^3 \, {\rm s^{-1}} \\
k_{-2} & = & 1.27 \times 10^3 \, {\rm s^{-1}} \\
k_{-3} & = & 5.29 \times 10^7 \, {\rm M^{-1} s^{-1}} \\
k_{-4} & = & 8.66 \times 10^7 \, {\rm M^{-1} s^{-1}} \\
k_{-5} & = & 3.75 \times 10^2 \, {\rm s^{-1}} \\
k_{-6} & = & 1.59 \times 10^3 \, {\rm s^{-1}} \\
k_{-7} & = & 9.52 \times 10^6 \, {\rm M^{-1} s^{-1}} \\
k_{-8} & = & 1.40 \times 10^8 \, {\rm M^{-1} s^{-1}}
\end{eqnarray*}
\end{minipage}
\end{minipage}
\vspace{0.5cm}
\caption[Rate constants in the kinetic model \cite{borgmann75}.]{Rate
  constants in the kinetic model \cite{borgmann75}.} 
\label{fig:model}
\end{figure}

\SubSection{Numerics}
\label{methods:comp}

\begin{figure}[t] \centering
\includegraphics[width=3in]{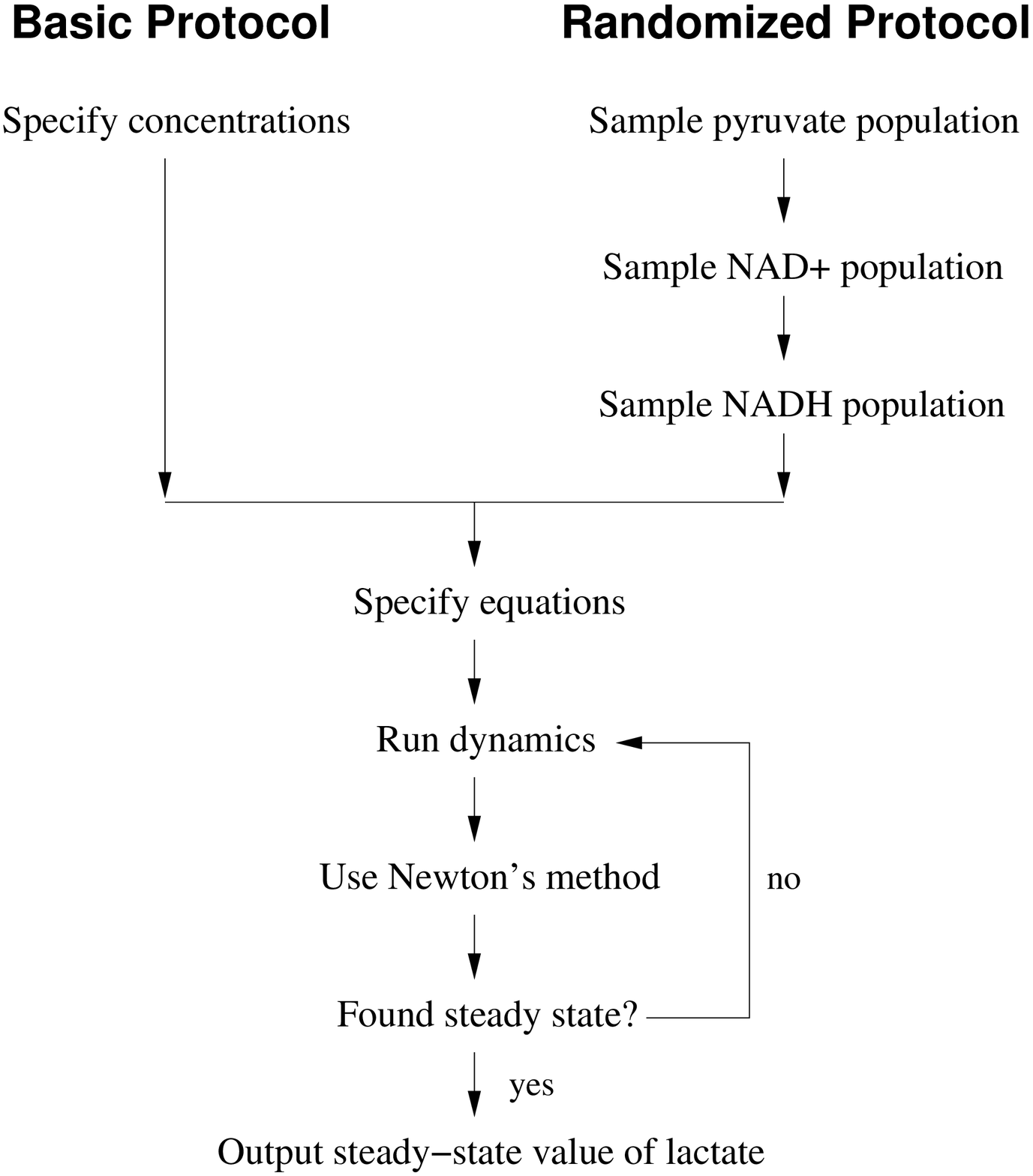}
\caption[Schematic of the computational protocol. The concentrations of
pyruvate, NAD+, and NADH are determined in one of two ways: either
they are specified (basic protocol) or they are randomly sampled
(randomized protocol). The total concentrations of LDH-H and LDH-M are
determined from the experimental conditions with both isoforms
present. We determine the steady-state concentration of lactate using
a hybrid approach. The model will relax to steady state if integrated
sufficiently long, but Newton's method accelerates the convergence to
steady state.]{Schematic of the computational protocol. The
  concentrations of pyruvate, NAD+, and NADH are determined in one of
  two ways: either they are specified (basic protocol) or they are
  randomly sampled (randomized protocol). The total concentrations of
  LDH-H and LDH-M are determined from the experimental conditions with
  both isoforms present. We determine the steady-state concentration
  of lactate using a hybrid approach. The model will relax to steady
  state if integrated sufficiently long, but Newton's method
  accelerates the convergence to steady state.}
\label{fig:workflow}
\end{figure}

A numerical approach was used to determine the steady-state
relationship between the metabolite concentrations, the metabolic
flux, and the isoform concentrations with both isoforms present
(figure \ref{fig:workflow}). Equations \eqref{m1}--\eqref{m3} can be
integrated numerically, allowing the system to approach a steady state
from any initial condition.  However, we specified the steady-state
concentrations of NAD+, NADH, and pyruvate, and determined the
corresponding steady-state concentration of lactate (equations
(\ref{eq1a}), (\ref{eq1c}), and (\ref{eq1d}) were eliminated from the
model).

The differential equations were integrated using a
Runge-Kutta-Fehlberg fifth-order method with adaptive stepping
\cite{press92}.  Newton's method was used to accelerate convergence to
the steady state.  If the solution to the constrained differential
equations was close to a steady state, Newton's method quickly
converged to it \cite{press92}.  If Newton's method started too far
from the steady state it rapidly diverged.  When this happened, the
constrained equations were integrated again (continuing from the last
solution) for a specified time interval.  Integrating for a longer
time allowed the system to approach closer to the steady state before
Newton's method was tried again. This hybrid approach worked well for
all the conditions we studied.

We note that Newton's method does not follow the dynamics of the
system and does not find the correct steady state unless the equations
are additionally constrained. In particular, it is necessary to
explicitly enforce the conservation of the total enzyme concentrations
(see below).

\Section{Results}
\noindent

\SubSection{Single-isoform results}

We examined the steady-state production of lactate by a single LDH
isoform. Under typical cellular conditions our model predicts that
LDH-H produces more lactate than LDH-M. This result is surprising
because many authors state that LDH-M produces lactate more efficiently
than LDH-H \cite{boyer63,stambaugh66,boyer75,voet04}. We explain the
reason for this difference, which
results from different model assumptions, and argue that our analysis
is more experimentally relevant.

\begin{figure}[h]\centering
\includegraphics[width=5in]{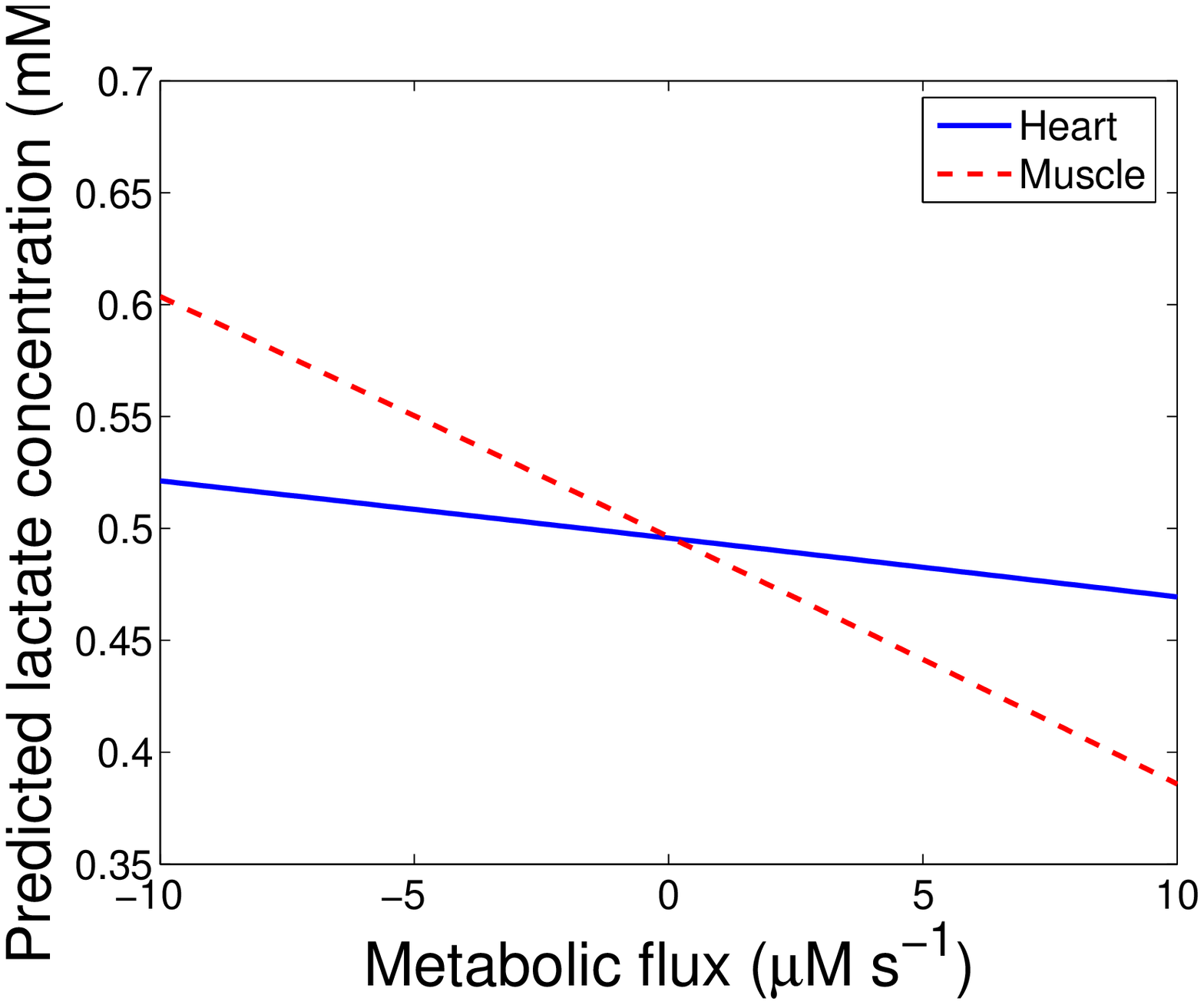}
\caption[The predicted concentration of lactate as a function of the
metabolic flux $\alpha$. In the model, a non-equilibrium steady state
is possible when pyruvate and NADH are added to the system at a
constant rate $\alpha$, and lactate and NAD+ are removed at the same
constant rate.  These terms in the model represent the production of
pyruvate (and consumption of lactate) in other chemical reactions or
transport into/out of the cell. As the metabolic flux $\alpha$ is
varied, the steady-state lactate concentration predicted by the model
changes. Note that $\alpha>0$ means that lactate is being removed from
the system. In each of the curves shown, only one of the LDH isoforms
is present. The concentrations of the metabolites and the total amount
of each enzyme are held constant (pyruvate, 99.4 \micromol; NAD+, 0.5
\millimol; NADH, 0.97 \micromol; LDH, 3.43 \micromol). The sign of the
metabolic flux determines which of the two isoforms favors the
production of lactate.  This conclusion differs from previous analyses
of this reaction, which were performed under Michaelis-Menten
conditions (see text). The experimental verification of this
prediction is future work.]{The predicted concentration of lactate as
  a function of the metabolic flux $\alpha$. In the model, a
  non-equilibrium steady state is possible when pyruvate and NADH are
  added to the system at a constant rate $\alpha$, and lactate and
  NAD+ are removed at the same constant rate.  These terms in the
  model represent the production of pyruvate (and consumption of
  lactate) in other chemical reactions or transport into/out of the
  cell. As the metabolic flux $\alpha$ is varied, the steady-state
  lactate concentration predicted by the model changes. Note that
  $\alpha>0$ means that lactate is being removed from the system. In
  each of the curves shown, only one of the LDH isoforms is present.
  The concentrations of the metabolites and the total amount of each
  enzyme are held constant (pyruvate, 99.4 \micromol; NAD+, 0.5
  \millimol; NADH, 0.97 \micromol; LDH, 3.43 \micromol). The sign of
  the metabolic flux determines which of the two isoforms favors the
  production of lactate.  This conclusion differs from previous
  analyses of this reaction, which were performed under
  Michaelis-Menten conditions (see text). The experimental
  verification of this prediction is future work.}
\label{fig:analytic}
\end{figure}

\SubSubSection{Steady-state lactate production by a single
isoform}

Here we are interested in the behavior of the model at steady state.
All concentrations are assumed to be steady-state values unless
otherwise stated. Suppose that the concentrations of pyruvate, NAD+,
and NADH are known. Furthermore, suppose only one isoform of LDH is
present and its total concentration is known. We can then derive an
equation that describes the concentration of lactate as a function of
the metabolic flux $\alpha$. Recall that when $\alpha>0$ lactate and
NAD+ are removed from the system and pyruvate and NADH are added to
the system.
 
The total concentration of either enzyme isoform is constant. This can
be shown for LDH-H by adding equations (\ref{eq2a})--(\ref{eq2d}),
\begin{equation}
e'_{1} + e'_{2} + e'_{3} + e'_{4} = 0,
\end{equation}
and integrating to obtain
\begin{equation}
e_{1} + e_{2} + e_{3} + e_{4} = e_{0}.
\end{equation}
Here $e_{0}$ is the total concentration of the heart isoform. This
relation always holds and is not particular to the steady state. The
same analysis applies to the muscle isoform:
\begin{equation}
f_{1} + f_{2} + f_{3} + f_{4} = f_{0},
\end{equation}
where $f_0$ is the total concentration of the muscle isoform.

At steady state the rates of change of all variables are zero, so the
derivatives in equations \eqref{eq1a}--\eqref{eq1d} are zero. This
implies that $v_{1} = v_{2} = v_{3} = v_{4} = -\alpha$. There are five
unknown quantities in the model: the four reaction intermediates
involving LDH-H, $e_{1}$, $e_{2}$, $e_{3}$, and $e_{4}$, and the
lactate concentration, $x_{2}$. (The other metabolite concentrations
and the metabolic flux are assumed known). The five equations required
for a unique solution are provided by the heart isoform conservation
law and the definitions of the reaction rates:
\begin{subequations}
\label{eq4}
\begin{eqnarray}
e_{1} + e_{2} + e_{3} + e_{4} & = & e_{0}, \label{eq4a} \\
k_{1} x_{1} e_{1} - k_{-1} e_{2} & = & -\alpha, \label{eq4b} \\
k_{2} x_{2} e_{2} - k_{-2} e_{3} & = & -\alpha, \label{eq4c} \\
k_{3} e_{3} - k_{-3} y_{1} e_{4} & = & -\alpha, \label{eq4d} \\
k_{4} e_{4} - k_{-4} y_{2} e_{1} & = & -\alpha. \label{eq4e}
\end{eqnarray}
\end{subequations}
Equations \eqref{eq4a}, \eqref{eq4b}, \eqref{eq4d}, and \eqref{eq4e}
form a linear system that determines $e_{1}$, $e_{2}$, $e_{3}$, and
$e_{4}$. The solutions for $e_{2}$ and $e_{3}$ can then be substituted
into \eqref{eq4c} to determine $x_{2}$. The result is
\begin{equation}
x_{2} = \frac{a_{0} e_{0} - a_{1} \alpha} {b_{0} e_{0} +
b_{1}
\alpha},
\label{eqn:analytic}
\end{equation}
where
\begin{eqnarray*}
a_{0} & = & k_{-1} k_{-2} k_{-3} k_{-4} y_{1} y_{2}, \\
a_{1} & = & k_{-1} k_{-2} k_{4} + k_{-1} k_{3} k_{4} + k_{-2} k_{1} k_{4} x_{1}+
 k_{1} k_{3} k_{4} x_{1} + k_{-1} k_{-2} k_{-3} y_{1} \\ 
&& + k_{-2} k_{-3} k_{1} x_{1} y_{1}+ k_{-1} k_{-2} k_{-4} y_{2} + 
 k_{-1} k_{-4} k_{3} y_{2} + k_{-1} k_{-3} k_{-4} y_{1} y_{2}  \\
&& + k_{-2} k_{-3} k_{-4} y_{1} y_{2}, \\
b_{0} & = & k_{1} k_{2} k_{3} k_{4} x_{1}, \\
b_{1} & = & k_{2} k_{3} k_{4} + k_{1} k_{2} k_{3} x_{1} +
k_{1} k_{2} k_{4} x_{1} + k_{-3} k_{1} k_{2} x_{1} y_{1} + k_{-4} k_{2} k_{3}
y_{2} \\
&& + k_{-3} k_{-4} k_{2} y_{1} y_{2}.
\end{eqnarray*}
Note that when $\alpha = 0$ we recover the equilibrium relationship
\begin{equation}
\frac{x_{1} x_{2}}{y_{1} y_{2}} = \frac{k_{-1} k_{-2}
k_{-3} k_{-4}}{k_{1} k_{2} k_{3} k_{4}} = K_{eq},
\end{equation}
where $K_{eq}$ is the equilibrium constant. In other
words, when
the metabolic flux is zero the system approaches an
equilibrium steady state, which is independent of the
total enzyme concentration.
\par

The concentration of lactate is a function of the metabolic flux,
$\alpha$, as shown in figure \ref{fig:analytic}. When the metabolic
flux is positive (lactate is being removed from the system) LDH-H
produces a higher steady-state lactate concentration than an equal
concentration of LDH-M. In our calculations we assumed concentrations
of NADH, NAD+, and pyruvate equal to 0.97 \micromol, 0.5 \millimol,
and 99.4 \micromol~respectively. However, this qualitative
trend---higher steady-state lactate concentration produced by
LDH-H---remains unchanged as long as the pyruvate concentration lies
between 40 \micromol~ and 2 \millimol.  Outside of this range the
steady-state solution is infeasible (at steady state one or more of
the metabolites has a concentration less than zero) or LDH-M produces
more lactate than LDH-H. This interval was found by varying NADH and
NAD+ independently over a wide range of concentrations (0.1 \micromol
-- 10.0 \millimol).  Previous measurements of cellular pyruvate
concentration have found a range 0.08 -- 0.3 \millimol, although the
concentration of pyruvate has been found as high as 0.7 \millimol~in
skeletal muscle under tetanic conditions
\cite{stambaugh66,boyer75,tilton91,lambeth2002b}.  Therefore, we
predict that in most cells LDH-H produces a higher steady-state
lactate concentration than an equal amount of LDH-M.
\par


Our result implies that LDH-H produces a greater concentration of
lactate in the cell then LDH-M does when the metabolic flux is
positive. Previous work on the kinetics of homolactic fermentation has
arrived at the opposite conclusion. The difference between our work
and previous studies \cite{stambaugh66} is that we have focused on the
steady state of the reaction. Previous analyses have adopted a
Michaelis-Menten approach, examining the behavior of the reaction far
from steady state and under \textit{in vitro} conditions. In
Michaelis-Menten theory the only possible steady state is the
equilibrium state. Our model includes equilibrium as a special case
when the metabolic flux is zero (see figure~\ref{fig:analytic}).
However, when the metabolic flux is nonzero the concentrations of the
enzymes and the reaction mechanism contribute to determining the
steady state.  Thus at steady state LDH-H results in more lactate than
LDH-M does when the metabolic flux is positive.  This relationship is
reversed when the metabolic flux is negative.

\SubSection{Predicted change in lactate concentration after MAP kinase
  activation} 

Here we predict how activating the MAP kinase pathway affects the
steady-state lactate concentration. Treating K562 cells with PMA
activates the MAP kinase pathway. Conversely, U0126 is a downstream
inhibitor of MAP kinase signaling which acts on MKK1 and
MKK2~\cite{gross00}. The array data show changes in the expression of
both LDH isoforms after activation of the MAP kinase pathway (figure
\ref{fig:affy_hist}). We used the relative abundances from our array
data to model three conditions: ({\it i}) control (cells are
untreated), ({\it ii}) MAP kinase active (PMA treatment), and ({\it
  iii}) MAP kinase partially suppressed (PMA+U0126 treatment).

We used the full model (figure \ref{fig:schematic}), with both
isoforms present. The array data determined the relative concentration
of each LDH isoform. The concentrations of the metabolites and enzymes
in homolactic fermentation were not measured in K562 cells. However,
metabolite and enzyme concentrations are available for muscle cells
and we used these data as reference values in our model. K562 cells
are unlikely to have an identical metabolic state to muscle cells, so
we performed a robustness analysis to determine how our results depend
on the reference concentrations used. Metabolite concentrations were
sampled over two orders of magnitude about the reference values. We
verified that our results remain qualitatively unchanged over this
range.  We used a total LDH concentration of 3.43
\micromol~\cite{mulquiney03} in the control condition and assumed that
the mRNA expression data directly predict protein concentrations.
Therefore, in the control model the total concentrations of LDH-H and
LDH-M were 1.98 \micromol~and 1.45 \micromol.  For the PMA-treatment
condition, the total concentrations of LDH-H and LDH-M were 1.83
\micromol~and 0.90 \micromol, while for the PMA+U0126-treatment
condition, the total concentrations of LDH-H and LDH-M were 1.27
\micromol~and 1.37 \micromol. The steady-state concentration of
lactate was determined numerically as described in Methods, with the
steady-state concentrations of NAD+, NADH, and pyruvate set to 0.5
\millimol, 0.97 \micromol, and 99.4 \micromol, respectively, and a
metabolic flux of 10.0 \micromol s$^{-1}$ \cite{lambeth2002b}.

\begin{figure}[t]\centering
\includegraphics[width=5in]{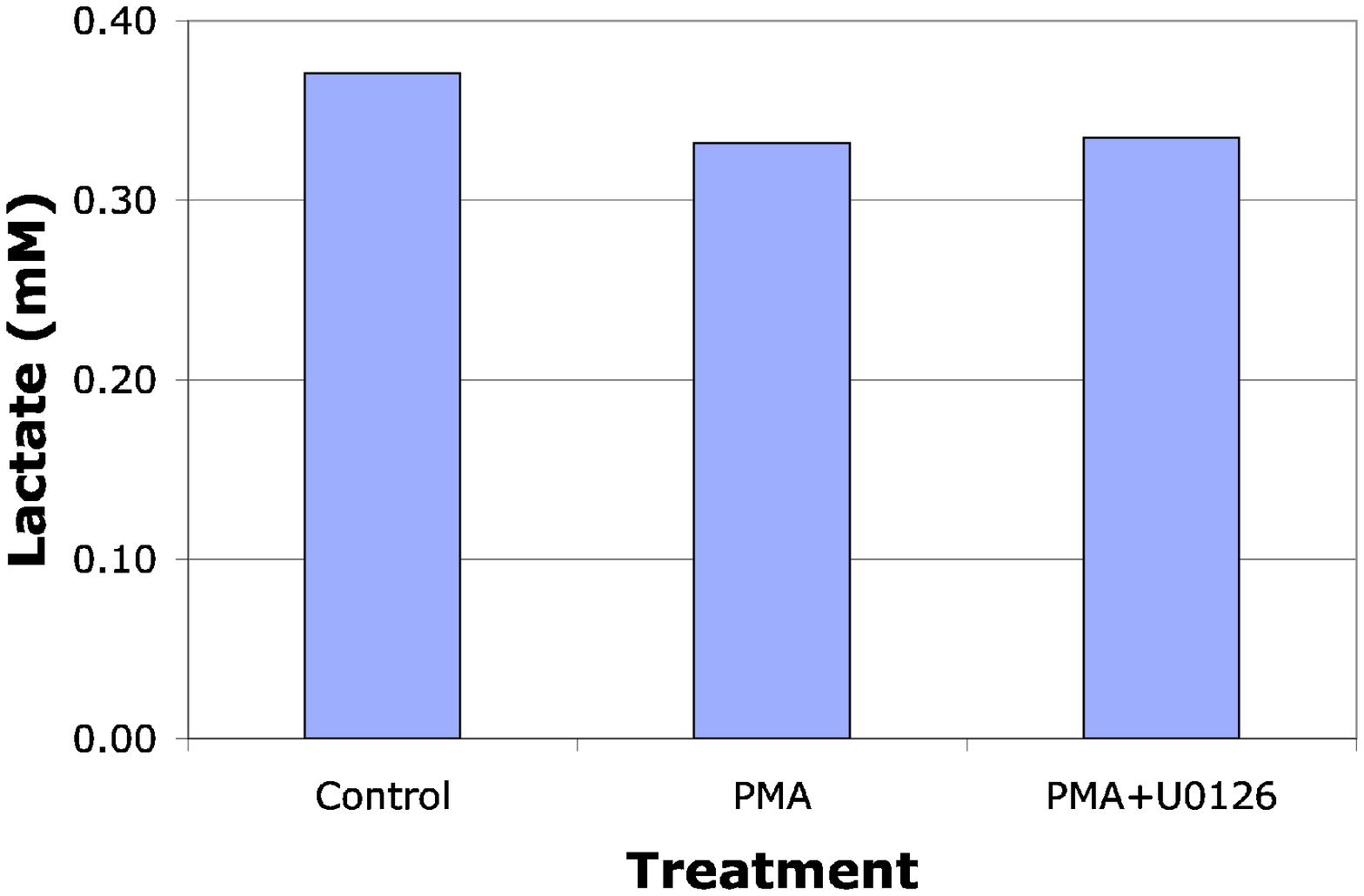}
\caption[Predicted concentration of lactate for three
conditions which simulate control, treatment with PMA, and treatment
with PMA+U0126.  The three conditions differ in the activity of the
MAP kinase pathway.  PMA treatment (which activates MAP kinase
signaling) results in a small but significant decrease in predicted
lactate concentration of 10.5\%. Treatment with PMA+U0126 slightly
increases the lactate level relative to the PMA treatment. The similar
lactate levels for PMA and PMA+U0126 are surprising, because the LDH
H:M ratio is 1.35 for the PMA treatment and 0.85 for the PMA+U0126
treatment.  The concentrations of pyruvate, NAD+, and NADH are 99.4
\micromol, 0.5 \millimol, and 0.97 \micromol.  The metabolic flux
$\alpha$ is 10.0 \micromol~ s$^{-1}$.  ]{Predicted concentration of
  lactate for three conditions which simulate control, treatment with
  PMA, and treatment with PMA+U0126.  The three conditions differ in
  the activity of the MAP kinase pathway.  PMA treatment (which
  activates MAP kinase signaling) results in a small but significant
  decrease in predicted lactate concentration of 10.5\%. Treatment
  with PMA+U0126 slightly increases the lactate level relative to the
  PMA treatment. The similar lactate levels for PMA and PMA+U0126 are
  surprising, because the LDH H:M ratio is 1.35 for the PMA treatment
  and 0.85 for the PMA+U0126 treatment.  The concentrations of
  pyruvate, NAD+, and NADH are 99.4 \micromol, 0.5 \millimol, and 0.97
  \micromol.  The metabolic flux $\alpha$ is 10.0 \micromol~
  s$^{-1}$.}
\label{fig:ref_bar}
\end{figure}

Our model predicts a decrease in lactate concentration for both the
PMA and PMA+U0126 conditions (figure \ref{fig:ref_bar}). The expression
data show that activating the MAP kinase pathway with PMA
down-regulates LDH-H and LDH-M (figure \ref{fig:affy_hist}). As a
result of these changes, the model predicts a decrease in the cellular
lactate level from 37.1 \micromol~to 33.2 \micromol. Treatment with
PMA+U0126 also down-regulates both LDH isoforms (figure
\ref{fig:affy_hist}), leading to a predicted lactate concentration of
33.5 \micromol.

\SubSubSection{Robustness analysis}
\label{robustness}

Here we demonstrate the robustness of the model predictions. The
experimentally observed changes in LDH-isoform mRNA predict a decrease
in lactate concentration after PMA or PMA+U0126 treatment, independent
of the precise values of metabolite concentrations used in the model.
The values of NADH, NAD+, and pyruvate concentrations can vary among
different cell types and growth conditions. While the quantitative
predictions depend on the values of the metabolite concentrations, the
qualitative result is robust.
 
We used each metabolite reference value as the mean of a sampled
distribution.  We chose a lognormal distribution to describe the
population of metabolite concentrations because samples vary by large
fractional amounts \cite{boyer75}.  The standard deviation is one
order of magnitude: approximately 68\% of the samples are within
0.1--10.0 times the mean concentration.  We note that the lognormal
distribution is nonnegative, guaranteeing nonnegative concentrations.
 
\begin{figure}[t]\centering
\includegraphics[width=5in]{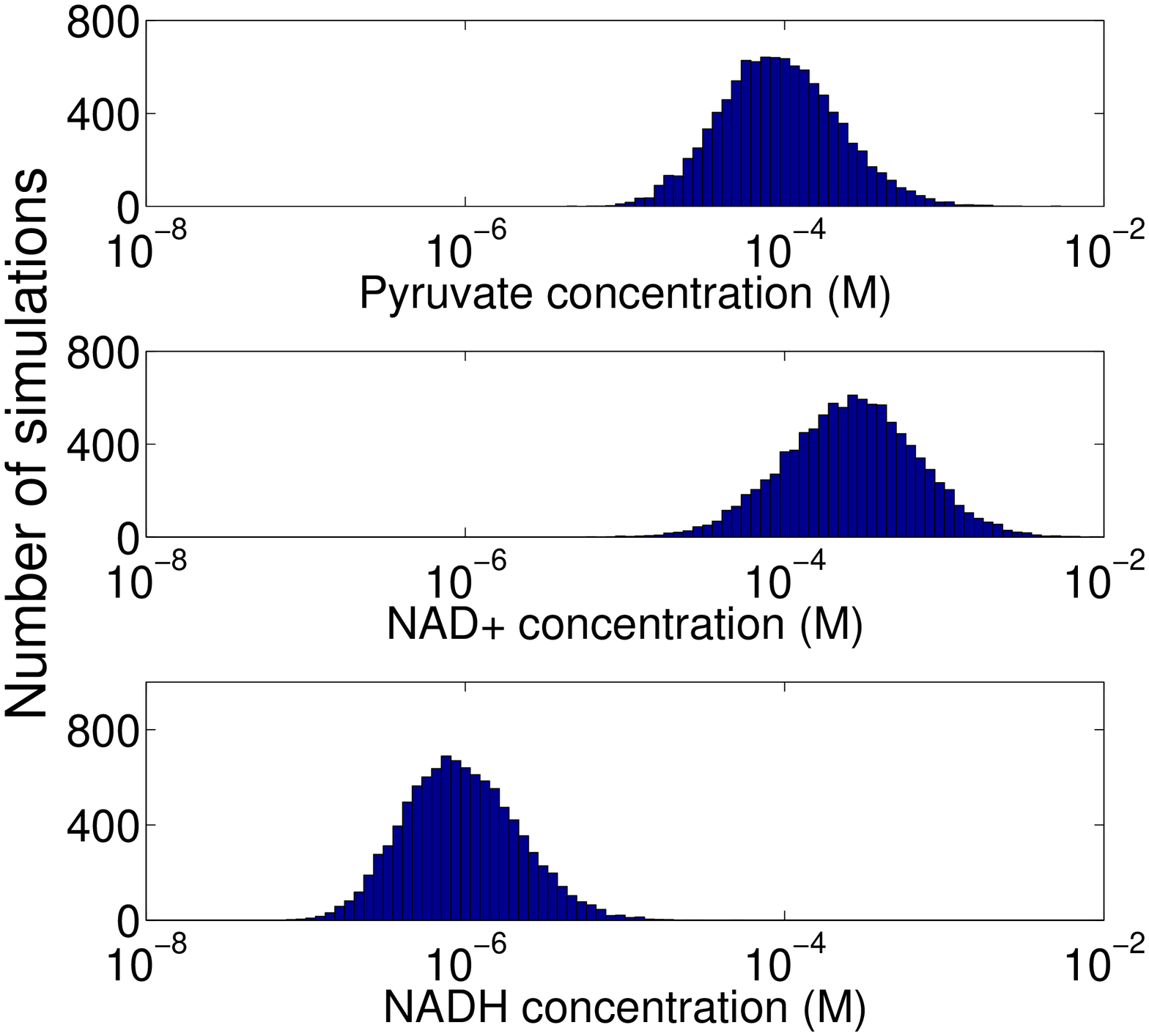}
\caption[Sampled metabolite concentration values. The distribution of
each concentration is lognormal (see text).  The mean of each
distribution coincides with the reference concentration of that
metabolite, and the standard deviation is one order of
magnitude.]{Sampled metabolite concentration values. The distribution
  of each concentration is lognormal (see text).  The mean of each
  distribution coincides with the reference concentration of that
  metabolite, and the standard deviation is one order of magnitude.}
\label{fig:metab_hist}
\end{figure}

We examined $10^4$ randomly sampled parameter sets (figure
\ref{fig:workflow}).  For each set of parameters, we calculated the
lactate concentration under three conditions: control, treatment with
PMA, and treatment with PMA+U0126. Due to our broad sampling of
parameter space, we found some parameter sets where the model cannot
reach a physically valid steady state (some parameters result in a
mathematical steady state with negative lactate concentration). Thus
we excluded from our analysis parameter sets that resulted in an
invalid steady state for {\em any} of the three conditions. This
exclusion did not significantly alter the lognormal distribution of
parameter values (figure \ref{fig:metab_hist}). We also checked the
consistency of the simulations by comparing the distribution of the
first 1,000 results with the distribution of the following 9,000
results. There was no statistically significant difference between
these two distributions by the Kolmogorov-Smirnov test ($p < 0.01$)
\cite{press92}.

\begin{figure}[t]\centering
\includegraphics[width=3in]{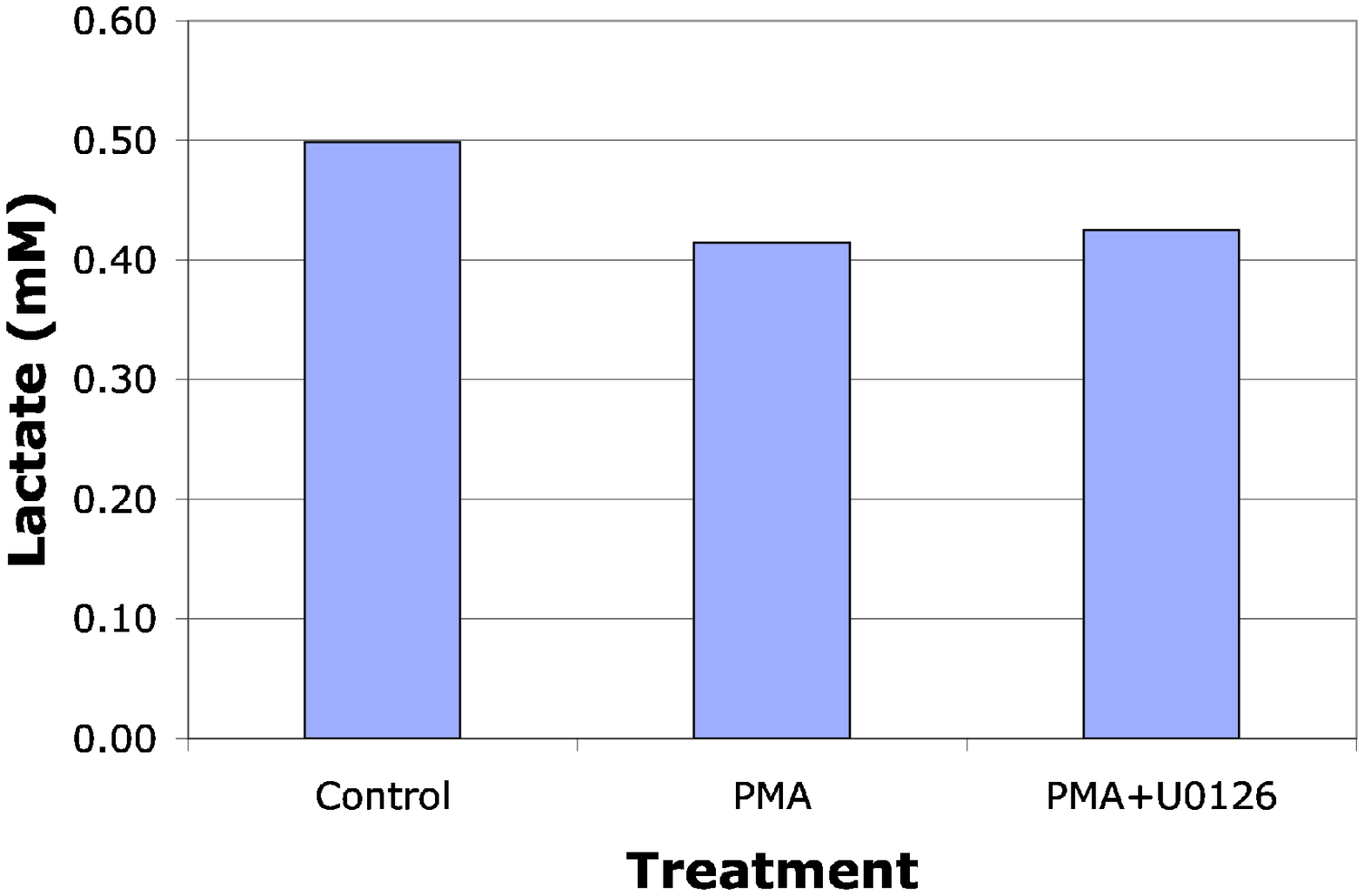}
\includegraphics[width=3in]{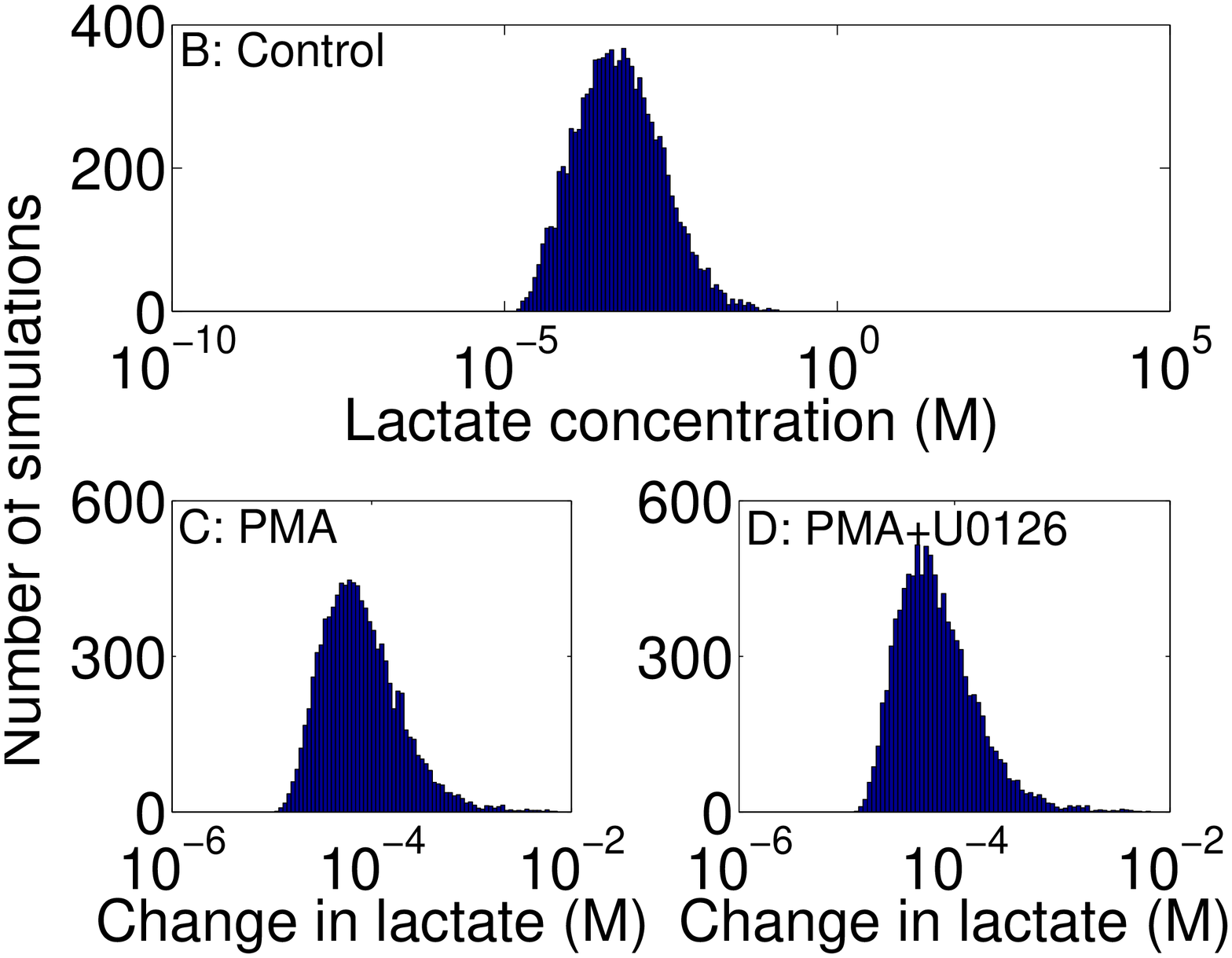}
\caption[Treating K562 cells with PMA or PMA+U0126 is predicted to
result in a similar concentration of lactate. (A) The median
concentration of lactate under the three conditions examined in the
robustness analysis (control, PMA, and PMA+U0126). This result is
qualitatively similar to the result from the simulations using only
the reference concentrations of pyruvate, NAD+, and NADH (figure
\ref{fig:ref_bar}). Treatment with PMA or PMA+U0126 decreases the
concentration of lactate. (B) Histogram of the predicted lactate
concentration for the control simulation. This panel illustrates the
distribution of the lactate concentration found in the robustness
analysis. The median of this histogram is the value of the control
treatment shown in A. (C) Histogram of the difference between the
lactate concentration predicted for the control and PMA treatment
conditions. All of the differences are positive, which means that PMA
treatment is predicted to decrease the steady-state lactate
concentration. After PMA treatment, the LDH isoform ratio increases
from 1.02 to 1.35. (D) Histogram of the difference between the lactate
concentration predicted for the control and PMA+U0126 treatment
conditions. All of the differences are positive, which means that
PMA+U0126 treatment is predicted to decrease the steady-state lactate
concentration. After PMA+U0126 treatment, the LDH isoform ratio
decreases from 1.02 to 0.85. ]{Treating K562 cells with PMA or
  PMA+U0126 is predicted to result in a similar concentration of
  lactate. (A) The median concentration of lactate under the three
  conditions examined in the robustness analysis (control, PMA, and
  PMA+U0126). This result is qualitatively similar to the result from
  the simulations using only the reference concentrations of pyruvate,
  NAD+, and NADH (figure \ref{fig:ref_bar}). Treatment with PMA or
  PMA+U0126 decreases the concentration of lactate. (B) Histogram of
  the predicted lactate concentration for the control simulation. This
  panel illustrates the distribution of the lactate concentration
  found in the robustness analysis. The median of this histogram is
  the value of the control treatment shown in A. (C) Histogram of the
  difference between the lactate concentration predicted for the
  control and PMA treatment conditions. All of the differences are
  positive, which means that PMA treatment is predicted to decrease
  the steady-state lactate concentration. After PMA treatment, the LDH
  isoform ratio increases from 1.02 to 1.35. (D) Histogram of the
  difference between the lactate concentration predicted for the
  control and PMA+U0126 treatment conditions. All of the differences
  are positive, which means that PMA+U0126 treatment is predicted to
  decrease the steady-state lactate concentration. After PMA+U0126
  treatment, the LDH isoform ratio decreases from 1.02 to 0.85.}
\label{fig:robust_median}
\end{figure}

The median predicted lactate concentration from the randomly sampled
parameters (figure \ref{fig:robust_median}) follows the same trend
observed for the reference parameter set (figure \ref{fig:ref_bar}).
The predicted lactate concentration decreases by 16.9\% for the PMA
treament and by 14.8\% for the PMA+U0126 treatment, relative to
control.  The systematic difference between the three conditions can be
seen by comparing sets of simulations corresponding to the same
parameter values. The difference histograms (figure
\ref{fig:robust_median}) show the decrease in lactate concentration
from the control to the treated condition.

The predicted lactate levels for the PMA and PMA+U0126 conditions are
similar. This result was unexpected because the LDH H:M isoform ratios
were different (1.35 for PMA and 0.85 for PMA+U0126). The key
difference between the two treatments and the control is the decrease
in the total concentration of LDH, an observation that led us to study
the effects of isoform switching and abundance changes separately.

\SubSection{Isolating the effects of isoform switching and
  abundance change}

We studied the effects of independently varying the isoform ratio and
LDH abundance using the robustness protocol described above (figure
\ref{fig:workflow}). First, we obtained concentrations for pyruvate,
NAD+, and NADH by sampling their distributions. The histograms of
sampled values are similar to those shown in figure
\ref{fig:metab_hist} (data not shown). For each sampled parameter set
we performed three simulations: ({\it i}) control, where the total
concentration of LDH was 3.43 \micromol~and the isoform ratio 1.02,
({\it ii}) increased isoform ratio, where the isoform ratio was 1.35
and the total LDH concentration remained 3.43 \micromol, and ({\it
  iii}) decreased total concentration, where the isoform ratio was the
control value (1.02) and the total concentration was 2.47 \micromol.
These values were taken from the mRNA expression data for control and
PMA treatment. This analysis allowed us to isolate the effects of
changing the isoform ratio and the total concentration of LDH. In each
simulation, the steady-state concentration of lactate was determined
numerically as described in Methods.
\par

\begin{figure}[t]\centering
  \includegraphics[width=3in]{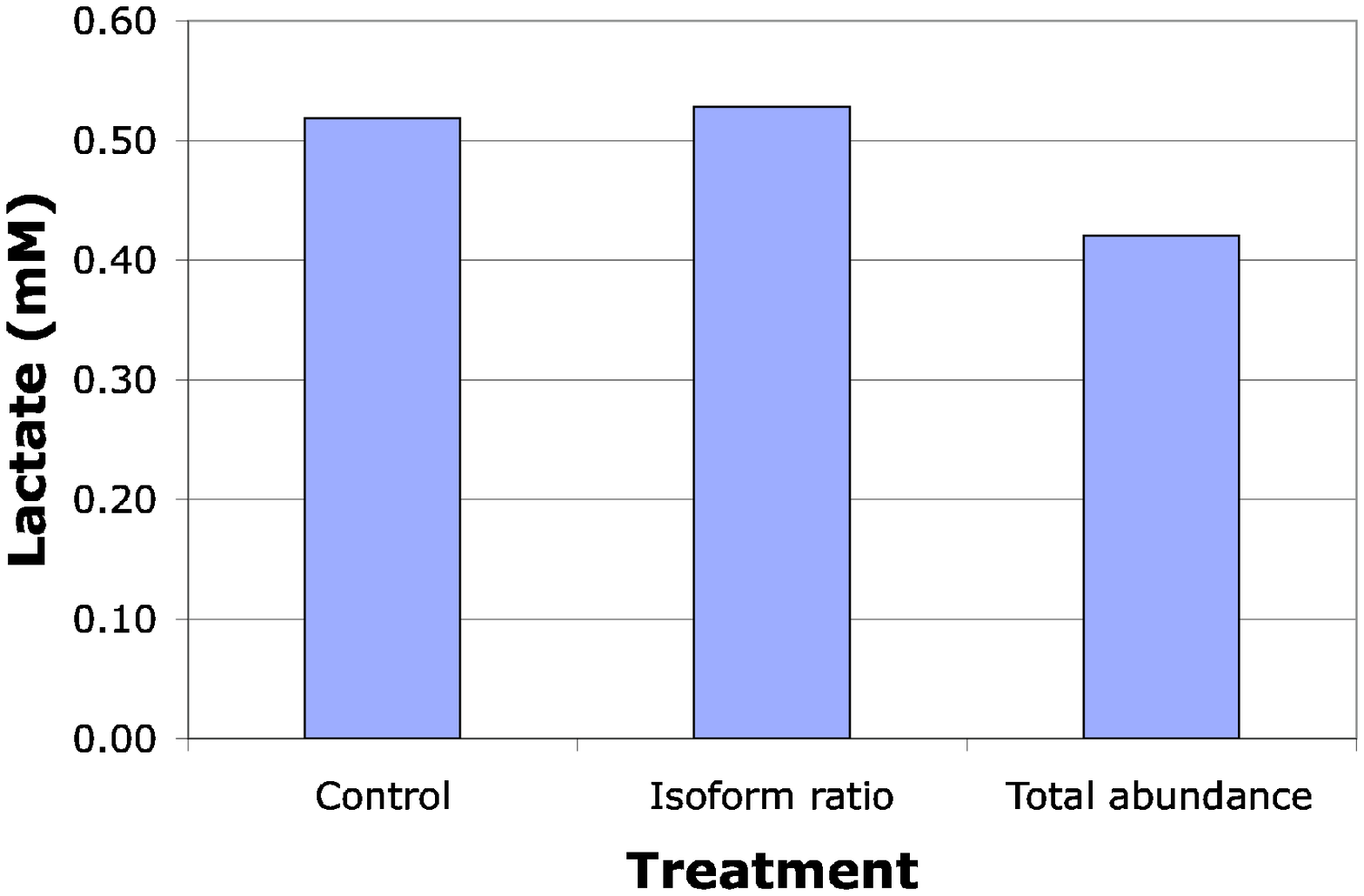}
  \includegraphics[width=3in]{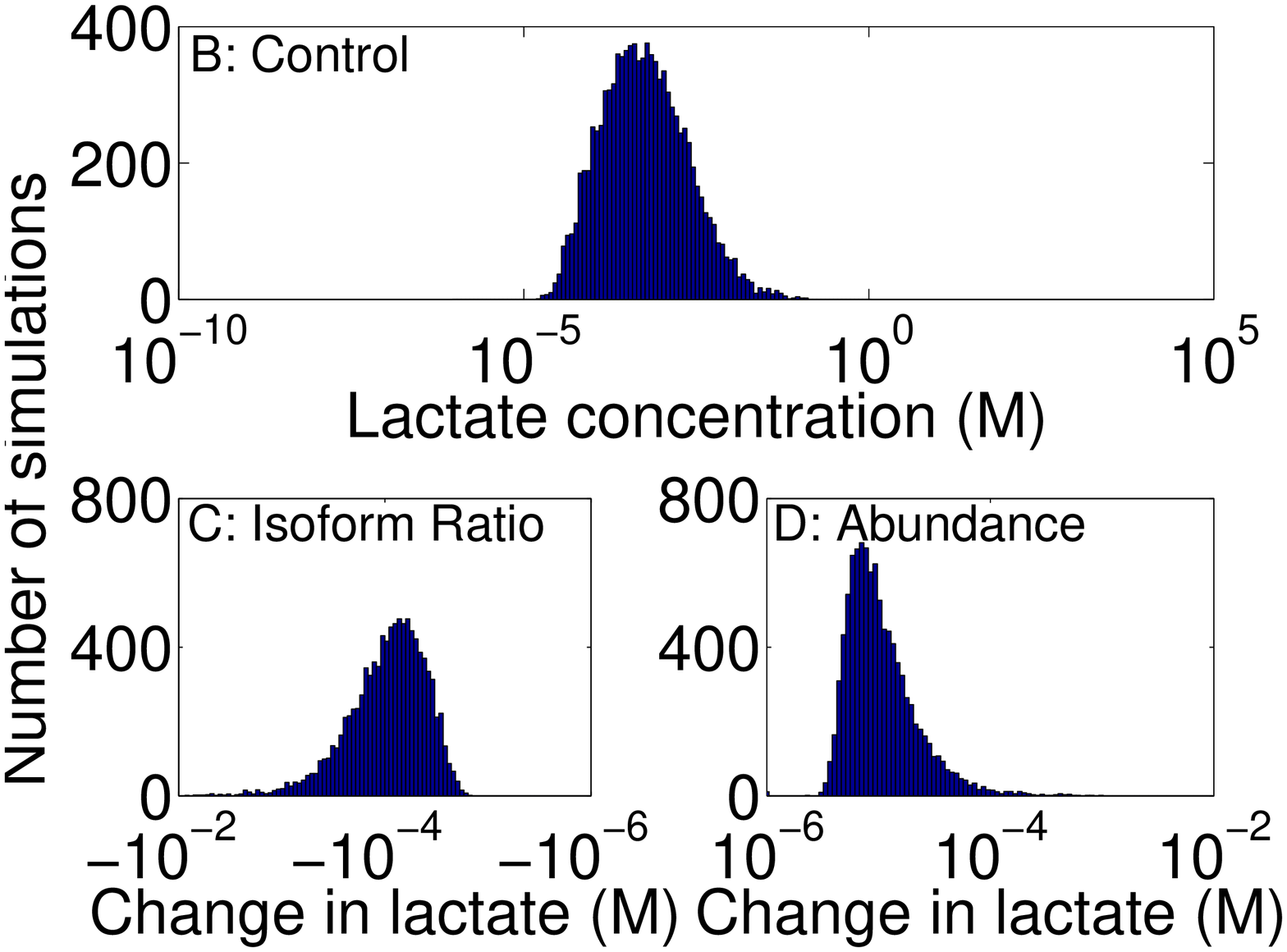}
\caption[The contrasting effects of increasing the isoform ratio and
decreasing the total concentration of LDH. (A) The median
concentration of lactate under the three conditions examined (control,
isoform-ratio increase, total LDH concentration decrease). When the
isoform ratio is increased and the total concentration of LDH is held
constant, the amount of lactate produced increases by a small amount.
This trend is consistent with the behavior of the individual isoforms:
LDH-H produces more lactate than LDH-M for $\alpha>0$.  When the
isoform ratio is held constant and the total concentration of LDH is
reduced, the amount of lactate produced decreases. (B) Histogram of
the predicted lactate concentration for the control simulation. This
panel illustrates the distribution of the lactate concentration found
in the robustness analysis. The median of this histogram is the value
of the control treatment shown in A. (C) Histogram of the difference
between the lactate concentration predicted for the isoform-ratio
increase and control conditions. Every set of metabolites resulted in
an increase in the predicted lactate concentration. (D) Histogram of
the difference between the lactate concentration predicted for the
total LDH concentration decrease and control conditions. Every set of
metabolites resulted in a decrease in the predicted lactate
concentration.  Note that the magnitude of the change in the lactate
concentration is approximately 10 times larger than the effect of
changing the isoform ratio (shown in C). The influence of the total
concentration of LDH on the production of lactate is much greater than
the influence of the isoform ratio.]{The contrasting effects of
  increasing the isoform ratio and decreasing the total concentration
  of LDH. (A) The median concentration of lactate under the three
  conditions examined (control, isoform-ratio increase, total LDH
  concentration decrease). When the isoform ratio is increased and the
  total concentration of LDH is held constant, the amount of lactate
  produced increases by a small amount.  This trend is consistent with
  the behavior of the individual isoforms: LDH-H produces more lactate
  than LDH-M for $\alpha>0$.  When the isoform ratio is held constant
  and the total concentration of LDH is reduced, the amount of lactate
  produced decreases. (B) Histogram of the predicted lactate
  concentration for the control simulation. This panel illustrates the
  distribution of the lactate concentration found in the robustness
  analysis. The median of this histogram is the value of the control
  treatment shown in A. (C) Histogram of the difference between the
  lactate concentration predicted for the isoform-ratio increase and
  control conditions. Every set of metabolites resulted in an increase
  in the predicted lactate concentration. (D) Histogram of the
  difference between the lactate concentration predicted for the total
  LDH concentration decrease and control conditions. Every set of
  metabolites resulted in a decrease in the predicted lactate
  concentration.  Note that the magnitude of the change in the lactate
  concentration is approximately 10 times larger than the effect of
  changing the isoform ratio (shown in C). The influence of the total
  concentration of LDH on the production of lactate is much greater
  than the influence of the isoform ratio.}
\label{fig:isolation}
\end{figure}

Figure \ref{fig:isolation} summarizes the effect of increasing the
isoform ratio or decreasing the total concentration of LDH. When the
isoform ratio is increased, the lactate concentration increases, and
when the total concentration is decreased, the lactate concentration
decreases.  Increasing the isoform ratio increased the lactate
concentration for every parameter set sampled.  Conversely, decreasing
the total concentration consistently decreases the lactate
concentration. The effect of decreasing the total concentration is
greater than the effect of changing the isoform ratio, given the
experimentally observed abundance changes in the isoforms. For these
experimental conditions, we predict that the total abundance of LDH is
more important than the isoform ratio in determining the lactate
concentration.

\Section{Discussion}

Our mRNA expression data show that activating MAP kinase signaling
changes the ratio of lactate dehydrogenase isoforms in K562 cells. We
used a mathematical model of lactate production by LDH to calculate
the changes in steady-state lactate concentration which result from
changes in LDH concentration. We assume that changes in gene
expression predict changes in enzyme concentration.  We predict that
for the experimentally observed changes in LDH, the cellular lactate
concentration undergoes a small but significant decrease. The
robustness analysis demonstrates that our prediction holds for a wide
range of metabolite concentrations. Experiments which measure the
lactate concentration in K562 cells under different conditions
(control and MAP kinase signaling active) can directly test our
prediction.

It is often stated in the literature that the two main LDH
isoforms---LDH-H and LDH-M---promote different directions of the
homolactic fermentation reaction. The idea that LDH-M favors the
production of lactate and LDH-H favors the production of pyruvate is
sometimes used to explain experimental results \cite{baker97,segal94}.
This interpretation of the role of the isoforms was based on
\textit{in vitro} biochemistry under Michaelis-Menten conditions,
which do not necessarily apply \textit{in vivo}.  Homolactic
fermentation is influenced by external supply of and demand for the
reactants and products, which means the reaction is not isolated.  In
addition to assuming an isolated reaction, Michaelis-Menten theory
describes the behavior of a reaction in its initial stages.  However,
metabolic reactions in the cell are typically close to steady state.

In our analysis, we focus on the nonequilibrium steady state of the
reaction in the presence of a metabolic flux, which represents the
production/consumption of NADH, NAD+, lactate, and pyruvate by other
sources in the cell. Under typical cellular conditions, LDH-H produces
a higher steady-state lactate concentration than does LDH-M. We
therefore state that LDH-H favors the production of lactate more than
LDH-M does. However, this does not imply that LDH-M favors the
production of pyruvate. Both isoforms favor the production of lactate
when pyruvate is supplied to the reaction (positive metabolic flux).
If the metabolic flux is negative, LDH-M produces more lactate than
LDH-H.

We show that examining changes in the LDH isoform ratio alone leads to
incorrect predictions: changes in total abundance of the isoforms must
also be considered. The changes in LDH isoform ratio we observe lead
to relatively small predicted changes in the amount of lactate. The
changes in total concentration of LDH lead to a larger predicted
change in lactate concentration. Taking into account total
concentration changes, as well as changes in isoform ratios, is
essential for a full understanding of the system.

An important problem in systems biology is the integration of
information from disparate sources~\cite{kitano02a}.  We describe an
approach to metabolic modeling that incorporates three important
components: (\textit{i}) the use of global profiling data to identify
an interesting problem and to guide the quantitative formulation of
the model; (\textit{ii}) a kinetic model which describes the full
dynamics of the system; and (\textit{iii}) a robustness analysis to
support the conclusions. To our knowledge, no previous metabolic
modeling work has incorporated all of these elements. The testing of
our approach on this small problem is a pilot study for applying the
method to larger systems, where the approach can be even more
valuable.

In recent years genome-scale profiling has become common.  Significant
hurdles remain in the interpretation and use of these data: how can
information from profiling be integrated to advance our knowledge? In
this paper, we began with an intriguing connection found in the
experiments: activating MAP kinase signaling changed the expression of
LDH isoforms. We predicted changes in cellular lactate metabolism
based on the data. The careful analysis of our profiling data allowed
us to isolate interesting and new effects.  We emphasize that our
model is based on experimentally observed changes in LDH expression,
so our results are experimentally relevant.

The advantage of using a kinetic model is that we can describe the
full dynamics of the system, including time-dependent behavior.
Valuable information about dynamics can arise from this approach; for
example, the time scale required for the system to reach steady state
can be determined. However, use of kinetic models is more complicated
than other approaches that do not consider the full dynamics.  The
major challenge in kinetic modeling is that many parameters are
unknown. Therefore, robustness analysis is essential. The robustness
analysis demonstrates that our conclusions do not apply only for a
specific parameter set, but are true in general. This component of our
approach addresses the fundamental problem of unknown parameters in
kinetic modeling.

In the future we hope to apply this approach to larger systems.
Testing the method on a smaller system (such as the one described in
this paper) is an important step in the development of the method.
However, we note that our results illustrate the power of carefully
analyzing small systems---surprising results can be obtained through
studies of this type.

\Section{Acknowledgements}

This work was supported by NIGMS project number 1540281. MDB
acknowledges support from the Alfred P. Sloan foundation.

\singlespacing


\end{document}